\documentclass[conference]{IEEEtran}
\IEEEoverridecommandlockouts

\usepackage{cite}
\usepackage{amsmath,amssymb,amsfonts}
\usepackage{graphicx}
\usepackage{textcomp}
\usepackage{xcolor}
\def\BibTeX{{\rm B\kern-.05em{\sc i\kern-.025em b}\kern-.08em
    T\kern-.1667em\lower.7ex\hbox{E}\kern-.125emX}}

\usepackage{xcolor}

\usepackage{times}
\usepackage{latexsym}
\usepackage[T1]{fontenc}
\usepackage[utf8]{inputenc}
\usepackage{microtype}

\usepackage{mwe}
\usepackage{booktabs}
\usepackage{multicol, multirow}
\usepackage{makecell}
\usepackage{subcaption}
\usepackage{array}
\usepackage{arydshln}
\usepackage{listings}
\usepackage[bookmarks=false]{hyperref}
\usepackage[normalem]{ulem}
\usepackage{url}
\definecolor{lightergray}{gray}{0.95} 
\definecolor{lightgray}{gray}{0.7}    
\newcommand{\model}{SCRAG\xspace}
\definecolor{linkblue}{rgb}{0.02,0.22,0.45}

\title{SCRAG: Social Computing-Based Retrieval Augmented Generation for Community Response Forecasting in Social Media Environments}

\author{
\IEEEauthorblockN{
Dachun Sun,
You Lyu,
Jinning Li,
Yizhuo Chen,
Tianshi Wang,
Tomoyoshi Kimura,
Tarek Abdelzaher
}
\IEEEauthorblockA{
\textit{Department of Computer Science}\\
University of Illinois Urbana-Champaign, Urbana IL 61801, USA\\
\{dsun18, youlyu2, jinning4, yizhuoc, tianshi3, tkimura4, zaher\}@illinois.edu
}
}

\begin{document}

\maketitle
\begin{abstract}
This paper introduces \model, a prediction framework inspired by social computing, designed to forecast community responses to real or hypothetical social media posts. \model can be used by public relations specialists (e.g., to craft messaging in ways that avoid unintended misinterpretations) or public figures and influencers (e.g., to anticipate social responses), among other applications related to public sentiment prediction, crisis management, and social what-if analysis. While large language models (LLMs) have achieved remarkable success in generating coherent and contextually rich text, their reliance on static training data and susceptibility to hallucinations limit their effectiveness at response forecasting in dynamic social media environments. \model overcomes these challenges by integrating LLMs with a Retrieval-Augmented Generation (RAG) technique rooted in social computing. Specifically, our framework retrieves (i) historical responses from the target community to capture their ideological, semantic, and emotional makeup, and (ii) external knowledge from sources such as news articles to inject time-sensitive context. This information is then jointly used to forecast the responses of the target community to new posts or narratives.
Extensive experiments across six scenarios on the $\mathbb{X}$ platform (formerly Twitter), tested with various embedding models and LLMs, demonstrate over 10\% improvements on average in key evaluation metrics. A concrete example further shows its effectiveness in capturing diverse ideologies and nuances. Our work provides a social computing tool for applications where accurate and concrete insights into community responses are crucial.
\end{abstract}


\begin{figure*}
    \centering
    \includegraphics[width=0.86\textwidth]{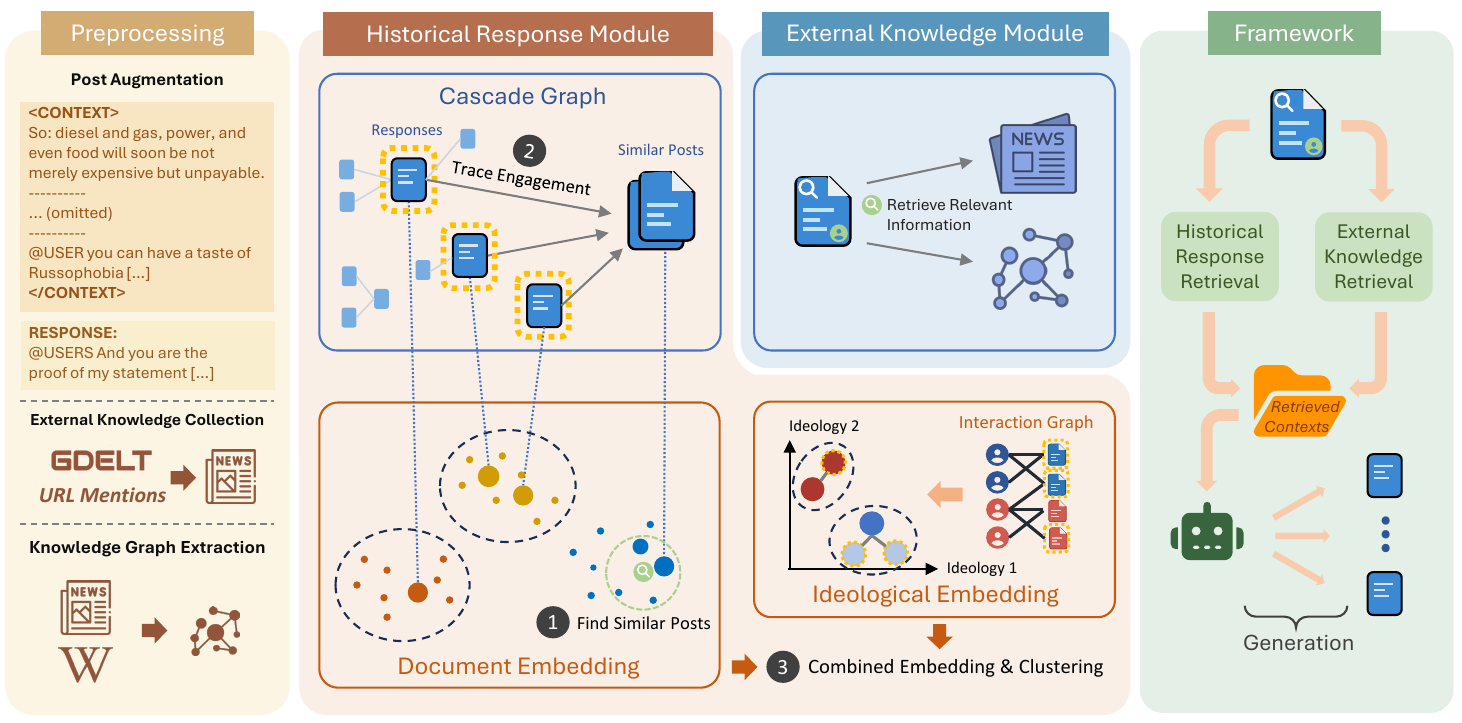}
    \caption{Framework architecture of \model.}
    \label{fig:arch}
    \vspace{-0.2in}
\end{figure*}

\section{Introduction}\label{sec:intro}

In today’s hyper-connected world, social media has evolved into a vibrant ecosystem where people engage in real-time dialogue, shape opinions publicly, and influence decision-making across various domains such as diplomacy, crisis response, and public relations. Public figures, public relation specialists, and social media analysts rely on these platforms to disseminate messages, manage reputations, and observe public reactions. Accurately forecasting community responses to potential social media narratives is thus helpful for preventing misunderstandings, navigating ideological differences, and proactively managing social dynamics.

Recent advances in large language models (LLMs), such as GPT~\cite{brown2020languagemodelsfewshotlearners}, have revolutionized natural language processing. LLMs excel in tasks that require understanding, summarizing, and generating texts, achieving remarkable success in applications ranging from question answering to conversation generation~\cite{wei2022emergent,achiam2023gpt,zhang2024benchmarking}. 
However, despite their capabilities, they encounter challenges when dealing with generation tasks related to dynamic and multifaceted social phenomena. In demanding social applications involving rapidly changing scenarios, such as crisis events, political developments, or emerging trends, LLMs often generate outdated and hallucinated responses that fail to reflect the nuances of evolving public sentiment.

To address these limitations, Retrieval-Augmented Generation (RAG)~\cite{lewis2020retrieval} has emerged as a promising approach to ground LLM outputs in new external knowledge. RAG combines the generative capabilities of LLMs with the recency and relevance of retrieved context, allowing them to access new information after their training, thereby improving their reliability when generating context-sensitive responses. This is particularly important in dynamic social media environments, where relevant and accurate context is crucial.

Building on these insights, we propose \model, a predictive framework that forecasts how communities will respond to real or hypothetical social media posts, reflecting diverse viewpoints and real-world conversational dynamics observed on platforms like $\mathbb{X}$ (formerly Twitter). Our framework combines LLMs with social computing-based RAG methods to enable the prediction of diverse, realistic, and contextually grounded responses. Specifically, we incorporate a community-aware historical response retriever to semantically and ideologically inform the forecasting. Furthermore, we retrieve relevant news articles and relations among entities for new scenario transfer and factual grounding.

Extensive experiments and a case study evaluate the emotional richness, ideological diversity, and contextual relevance of predicted responses (compared to real ones, withheld from the forecasting algorithm) using various metrics, with \model demonstrating superior performance in almost all cases. Unlike aggregated summarization approaches, these fine-grained forecasts support applications where understanding the precise expression of community reactions is important. Our contributions are summarized as follows:
\begin{itemize}
    \item We present a novel framework that integrates LLMs with social computing-based RAG techniques to forecast diverse, realistic, and contextually grounded responses.
    \item We demonstrate that our framework is highly modular and can admit various embedding models and LLMs, making it adaptable to different needs and resource availability.
    \item We evaluate our framework through extensive experiments on multiple datasets, highlighting its ability to produce accurate and rich predictions.
\end{itemize}

\noindent
The rest of this paper is organized as follows. 
Section~\ref{sec:definition} presents our problem statement. 
Section~\ref{sec:method} details our proposed framework, followed by experimental results in Section~\ref{sec:exp}. Section~\ref{sec:related} reviews related work in response generation, RAG techniques, and ideological embedding. Finally, Section~\ref{sec:conclusion} concludes the paper and outlines directions for future research.

\section{Problem Definition}\label{sec:definition}

This paper addresses the challenge of accurately predicting community responses to selected real or hypothetical social media posts. Our objective is to create a framework that automatically captures the spectrum of community beliefs while adapting to new external information, offering realistic expressions, representing ideological diversity, ensuring contextual relevance, and adapting to changing circumstances.

Let \( p_s \) denote a new (real or hypothetical) post on a social medium, and let $\mathcal{D}_p = \{ (p_i, \mathcal{R}_i) \mid i = 1, \dots, N_p \}$ be an extensive historical response dataset consisting of previous posts on various topics, where \( p_i \) is any post with responses, and \( \mathcal{R}_i \) represents the set of those responses (replies, quotes, or comments).
Also, let $\mathcal{D}_n = \{ d_i \mid i = 1, \dots, N_n \}$ represent a continuously updated news dataset. Given $\mathcal{D}_p$ and $\mathcal{D}_n$, the goal is to generate a set of realistic, diverse, and contextually grounded responses for post $p_s$.

\section{Methodology}\label{sec:method}

We introduce \model, a framework designed to predict community responses on social media via \textbf{S}ocial \textbf{C}omputing-based \textbf{R}etrieval-\textbf{A}ugmented \textbf{G}eneration. Figure~\ref{fig:arch} illustrates an overview of the proposed framework.
Our approach consists of two main modules inspired by social computing: (i) \emph{Community-aware retrieval of historical responses}, where past responses under a similar scenario are retrieved and clustered into semantically and ideologically coherent groups, providing community ideology identification and writing style references for response forecasting, and (ii) \emph{Sparse retrieval of external knowledge} for obtaining up-to-date relevant news articles and knowledge graph 
(KG) relations, grounding responses in appropriate contexts.

\subsection{\bf Preprocessing}

\subsubsection{\bf Historical Posts Augmentations}\label{1_doc_aug}
On social media, especially on short-text platforms like $\mathbb{X}$, a single post often cannot be fully understood without the surrounding conversational context. Therefore, for each post $p_i$ in the response chain, we trace its path back to the root (original) post and convert the trace into a text document. Each augmented post includes the root message and few immediate parent posts along the chain as context. This approach preserves important information and provides the necessary details to understand post $p_i$. 

\subsubsection{\bf External Knowledge Collection and KG Extraction}\label{1_kg_col_ext}
First, we collect mentions of news articles from social media posts and scrape their content to create the initial set $\mathcal{D}_n$. 
To maintain an up-to-date database for the prediction framework, we continuously gather additional news articles from the GDELT project~\cite{gdelt}, which serves as an external source for global events. As the news articles are collected, they undergo an LLM pipeline to extract entities and relations, forming parts of the knowledge graph $G_K$. To facilitate the retrieval of knowledge graph relation triplets, we convert nodes in $G_K$ along with their local neighborhoods into text representations, resulting in a set of documents referred to as $\mathcal{D}_g$. If a node has too many neighbors, the text representation is divided into chunks. If there are not enough neighboring nodes, we temporarily disregard the document until they become available.

\subsection{\bf Community-Aware Historical Response Retriever}
Given the new post $p_s$ on social media, we will retrieve historical responses to a similar context from the past and cluster them into ideologically and semantically distinct groups, establishing a foundation for predicting how communities will react to the new post. The purpose of this module is twofold: (i) to capture the ideological and semantic makeup for the forecast by automatically identifying different communities and summarizing their dominant beliefs from the data, and (ii) to prepare examples of the emotional and writing style characteristic of each community to predict realistic responses.

\subsubsection{\bf Constructing the Historical Response Database}\label{1_doc_db}
This retriever module depends on the historical response database, and we present the procedure to populate it. Let \(\mathcal{E}: \text{doc}\rightarrow\mathbb{R}^d\) be an embedding model that maps a document to a $d$-dimensional vector with an instruction shown in Figure~\ref{prompt:ours}, and let \(\mathcal{V}\) be a vector database that indexes based on Euclidean distances between vectors. We populate $\mathcal{V}$ with embeddings of augmented posts in $\mathcal{D}_p$ using $\mathcal{E}$.

Text embeddings effectively represent semantic information, which is important for retrieving similar posts. However, semantics do not definitively relate to ideologies or beliefs of communities. We can also derive ideological embeddings from these posts through the \textcolor{black}{user-post interaction bipartite graph $G_I=(\mathcal{U}\cup \mathcal{P}, E)$}, inherent to social networks. Ideological representation learning algorithms~\cite{li2022unsupervised,li2024large} create embeddings for users and posts, ensuring that those who share the same ideology are positioned close together in the latent space by minimizing the reconstruction loss of the graph structure:
\begin{equation}
\mathbb{P}\left((i\rightarrow j) \in E~|~\mathbf{u}_i,\mathbf{m}_j\right)=\sigma(\mathbf{u}_i^T\mathbf{m}_j)
\end{equation}
where $\mathbf{u}_i,\mathbf{m}_j\in \mathbb{R}^{d'}$ are the ideological embeddings of user $i$ and post $j$. 
We compute these embeddings of augmented posts for each topic separately and insert them into $\mathcal{V}$ with appropriate padding. For each topic, the ideological sides default to \emph{pro-} and \emph{anti-}. The following section will use these alongside text embeddings to define communities.

\subsubsection{\bf Historical Responses Retrieval}
Given a new augmented post $p_s$ and $\mathbf{e}_s = \mathcal{E}(p_s)$, we retrieve a set of $k_p+k_\Delta$ similar posts \(\mathcal{P}_{\text{candidate}}\) from the vector store \(\mathcal{V}\) ordered by their distances to $\mathbf{e}_s$, where $k_p$ and $k_\Delta$ are hyperparameters that need tuning based on the size of the historical database and the coverage of different ideologies and semantics. 
We further refine the candidate set by prompting an LLM agent to determine for each $p_i\in \mathcal{P}_{\text{candidate}}$ whether it relates to $p_s$ in a broader category or expresses similar stances toward these categories. The refined set $\mathcal{P}_{\text{similar}}$ is then traced to locate the historical responses. The output of this step is the union of all \(\mathcal{R}_i\), containing historical responses to posts $p_i \in \mathcal{P}_{\text{similar}}$ deemed similar to $p_s$, denoted as:
\begin{equation}
\mathcal{R}_\text{gather} = \cup_i\{\mathcal{R}_i | p_i \in \mathcal{P}_{\text{similar}} \}
\end{equation}

\subsubsection{\bf Community Discovery with Clustering}\label{3_cluster}
We attempt to discover communities computationally based on the text embeddings and ideological embeddings of the gathered responses. By combining them, the clustering algorithm gains more information, becoming semantically and ideologically aware. We confirmed the effectiveness of this approach through our experiments.
Since the dimension of the text embedding is much higher, we apply UMAP~\cite{mcinnes2018umap} first before concatenating it with the ideological embedding. 
We then apply HDBSCAN~\cite{mcinnes2017hdbscan}, a density-based clustering algorithm, to $\mathcal{R}_\text{gather}$ using the combined embeddings, forming $N_C$ clusters.
Responses identified as outliers by the clustering algorithm are split further into separate sets using ideological embedding to respect the ideological differences among the outliers. Ultimately, we have a set of discovered communities:
\begin{equation}
    \{C_1, C_2, \dots, C_{N_C}, (C_{O,1}, \dots, C_{O,d'})\}
\end{equation}

We select $k_c$ representative responses from each cluster for the generation module. To ensure diversity in the final selection, we employ a strategy similar to maximal marginal relevance (MMR)~\cite{carbonell1998use}, based on the responses' proximity to the cluster medoid and its semantic similarity to the selected responses.

\subsection{\bf Sparse Retrieval of External Knowledge}
In addition to historical responses, predicting community responses requires additional factual and time-sensitive knowledge, especially for emerging scenarios involving new events and entities absent from historical data. To address this, we employ a sparse retrieval module to fetch up-to-date external knowledge from news articles and the GDELT project~\cite{gdelt}, both considered external knowledge sources in this paper.

Sparse retrieval is used for its lower computational demands and the domain shifts between short social media queries and longer news articles, where semantic-based approaches might be less suitable. At the same time, pure term-based retrieval may be too restrictive for our application, making SPLADE~\cite{formal2021splade} an ideal component for this module because it combines term-based matching with query expansion, reducing limitations and enhancing retrieval performance.

\subsubsection{\bf News Articles and KG Relations Retrieval}
Documents in $\mathcal{D}_n$ and $\mathcal{D}_g$ are prepared in Section~\ref{1_kg_col_ext}. We embed them using the SPLADE model $\mathcal{E}'$ and insert them into separate vector databases that index based on sparse inner products between sparse vectors. Newly collected articles are also processed similarly on time. Given $p_s$ and $e_s'=\mathcal{E}'(p_s)$, we retrieve $k_n$ news article snippets and $k_g$ relation triplet chunks.

\subsection{\bf Generation Module}
\begin{figure}[htb]
\centering
\includegraphics[width=0.98\linewidth]{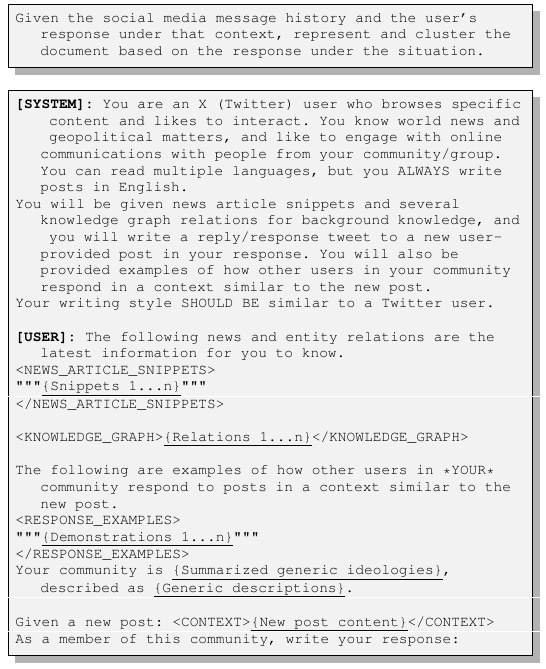}
\caption{Special instruction that focuses the embedding model on the response, and LLM prompt used by \model that has historical responses with ideologies and external knowledge.}
\label{prompt:ours}
\end{figure}

We prompt an LLM to forecast $M$ new responses that respect the identified communities and external knowledge. In Figure~\ref{prompt:ours}, we show \model's dedicated prompt that incorporates retrieved historical responses, ideological attributes of the community, relevant news articles, and relevant knowledge graph relation triplets. To also predict the potentially different activity level of each community, we heuristically allocate the quota of $M$ responses based on identified community sizes.
\begin{equation}
M_k \sim \frac{|C_k|}{\sum_{k=1}^{N_C} |C_k|} \cdot M
\end{equation}

For each cluster \(C_k\), we prompt the LLM to generate \(M_k\) responses.
To prevent the generation of highly repeated responses, our framework can prompt the LLM sequentially with additional instructions to discourage similar expressions or prompt in parallel while randomizing the sampling seed during decoding.

\begin{table}[hbt]
\caption{Statistics of cascade and interaction graphs (after preprocessing) of the datasets that are included in the historical response database.}
\label{tbl:datasets}
\centering
\resizebox{\linewidth}{!}{
\begin{tabular}{cccccc}
\toprule
\multirow{2}{*}{Datasets} & \multicolumn{2}{c}{Cascade Graph} & \multicolumn{3}{c}{Interaction Graph} \\ \cmidrule(lr){2-3} \cmidrule(lr){4-6} 
 & \#Docs & \#Edges & \#Assertions & \#Users & \#Edges \\ \midrule
Russophobia & 92,431 & 69,117 & 58,840 & 80,075 & 177,550 \\
US2020 & 109,771 & 89,422 & 66,898 & 130,045 & 172,044 \\
COVID & 67,721 & 63,275 & 43,093 & 108,584 & 113,286 \\
UK & 137,418 & 96,047 & 291,012 & 307,539 & 1,014,606 \\
\bottomrule
\end{tabular}
}
\end{table}

\section{Experiments}\label{sec:exp}

\subsection{Experimental Setup}
In this section, we assess the performance of the \model framework under six real scenarios on the $\mathbb{X}$ platform. 
To evaluate the predictions, we set aside 30 selected testing posts with abundant real responses (30-60) for each scenario.
\model outperforms other baselines for all four metrics in most scenarios with various embedding and LLM models. 
A generation example will present directly predicted responses from our framework, and we also show that all components of \model are important through ablation studies.
We use public and self-collected datasets for four scenarios and preprocessed them into a historical response database according to Section~\ref{1_doc_aug} and \ref{1_doc_db}.
Test posts collected for the two other scenarios are employed to test the adaptability of our framework by evaluating the predictions using similar historical responses transferred and grounded by the external knowledge.
The code of \model is available at \textcolor{linkblue}{\url{https://github.com/dsun9/SCRAG}}.

\begin{table*}[h!]
\caption{Evaluation results of \model against two other baselines with different embedding models for each scenario. LLM is Llama3.3-70B, and we report averages for 30 test cases in each scenario. Bold entries are the best-performing. Arrows after metric names indicate whether higher or lower is better.}
\label{tbl:main_emb}
\centering
\resizebox{0.98\textwidth}{!}{%
\begin{tabular}{llcccccccccccc}
\toprule

\multirow{2}{*}{Embedding} & \multirow{2}{*}{Methods} & \multicolumn{6}{c}{Emotion JSD $\downarrow$} & \multicolumn{6}{c}{Cluster Matching (\%) $\uparrow$} \\ \cmidrule(lr){3-8} \cmidrule(lr){9-14} 
 &  & Russophobia & US2020 & COVID & UK & US2024 & Gaza & Rus (72) & US20 (82) & COV (87) & UK (60) & US24 (79) & Gaza (75) \\ \midrule
 
\multirow{3}{*}{VoyageAI} & Direct & 0.453 & 0.261 & 0.357 & 0.437 & 0.474 & 0.348 & 47.41 & 64.07 & \textbf{61.67} & 42.00 & 59.63 & 42.67 \\
 & Fewshot & 0.315 & 0.275 & 0.287 & 0.285 & 0.362 & 0.245 & 55.19 & 59.26 & 57.78 & 45.67 & 65.56 & 38.67 \\ 
 & \model & \textbf{0.223} & \textbf{0.214} & \textbf{0.265} & \textbf{0.223} & \textbf{0.283} & \textbf{0.212} & \textbf{55.56} & \textbf{67.04} & 58.33 & \textbf{48.67} & \textbf{68.89} & \textbf{47.67} \\ \midrule 
\multirow{3}{*}{OpenAI} & Direct & 0.436 & 0.266 & 0.356 & 0.443 & 0.479 & 0.362 & 45.56 & 64.81 & 60.56 & 45.67 & 57.78 & 41.67 \\
 & Fewshot & 0.268 & 0.269 & 0.341 & 0.281 & 0.419 & 0.261 & 54.44 & \textbf{68.89} & 58.89 & 46.00 & 64.81 & 35.00 \\ 
 & \model & \textbf{0.263} & \textbf{0.265} & \textbf{0.306} & \textbf{0.240} & \textbf{0.335} & \textbf{0.236} & \textbf{55.19} & 64.81 & \textbf{62.22} & \textbf{49.33} & \textbf{68.52} & \textbf{43.33} \\ \midrule 
\multirow{3}{*}{NV-Embed2} & Direct & 0.430 & 0.285 & 0.375 & 0.456 & 0.477 & 0.384 & 51.48 & 55.19 & 63.33 & 40.67 & 55.56 & 35.67 \\
 & Fewshot & 0.306 & 0.254 & 0.377 & 0.270 & 0.394 & 0.282 & 50.37 & 62.96 & \textbf{56.67} & 45.67 & 64.81 & 46.25 \\ 
 & \model & \textbf{0.273} & \textbf{0.246} & \textbf{0.321} & \textbf{0.261} & \textbf{0.343} & \textbf{0.248} & \textbf{53.70} & \textbf{71.85} & 52.67 & \textbf{52.33} & \textbf{68.15} & \textbf{49.17} \\ 

\midrule \midrule 

\multirow{2}{*}{Embedding} & \multirow{2}{*}{Methods} & \multicolumn{6}{c}{LLM Discrimination Score $\uparrow$} & \multicolumn{6}{c}{Cluster Coverage (\%) $\uparrow$} \\ \cmidrule(lr){3-8} \cmidrule(lr){9-14} 
 &  & Russophobia & US2020 & COVID & UK & US2024 & Gaza & Russophobia & US2020 & COVID & UK & US2024 & Gaza \\ \midrule
 
\multirow{3}{*}{VoyageAI} & Direct & 7.822 & 8.559 & 7.033 & 8.077 & 8.037 & 7.957 & 58.52 & 39.81 & 25.00 & 51.33 & 33.15 & 55.83 \\
 & Fewshot & 8.344 & 8.500 & 7.050 & 8.420 & \textbf{8.178} & 8.073 & 59.26 & 58.33 & 26.19 & 54.00 & 46.48 & 59.17 \\ 
 & \model & \textbf{8.552} & \textbf{8.730} & \textbf{7.350} & \textbf{8.967} & 8.163 & \textbf{8.140} & \textbf{77.78} & \textbf{63.89} & \textbf{37.50} & \textbf{66.50} & \textbf{60.19} & \textbf{62.50} \\ \midrule 
\multirow{3}{*}{OpenAI} & Direct & 7.863 & 8.407 & 7.039 & 7.960 & 8.019 & 8.107 & 62.04 & 45.37 & 25.00 & 47.33 & 38.33 & 52.50 \\
 & Fewshot & 8.385 & \textbf{8.567} & 6.678 & 8.587 & 7.889 & 8.003 & 62.22 & 47.22 & 33.33 & 63.17 & 40.56 & \textbf{64.17} \\ 
 & \model & \textbf{8.430} & 8.526 & \textbf{7.172} & \textbf{8.660} & \textbf{8.144} & \textbf{8.133} & \textbf{62.96} & \textbf{56.48} & \textbf{37.00} & \textbf{65.17} & \textbf{53.52} & 45.83 \\ \midrule 
\multirow{3}{*}{NV-Embed2} & Direct & 7.811 & 8.467 & 6.720 & 7.953 & 8.000 & 7.893 & 51.85 & 39.81 & 27.38 & 50.67 & 35.37 & 37.50 \\
 & Fewshot & 8.337 & 8.470 & 7.350 & 8.677 & 7.967 & 8.100 & 53.70 & 52.78 & 33.00 & 65.67 & 36.85 & 47.92 \\ 
 & \model & \textbf{8.422} & \textbf{8.504} & \textbf{7.707} & \textbf{8.693} & \textbf{8.081} & \textbf{8.171} & \textbf{64.81} & \textbf{52.96} & \textbf{35.00} & \textbf{66.50} & \textbf{47.96} & \textbf{54.17} \\ 
 
\bottomrule
\end{tabular}
}
\end{table*}
\begin{table*}[h!]
\caption{Evaluation results of \model against fewshot baseline with more LLMs (three models with increasing number of parameters and a commercial model) for each scenario. Embedding model is VoyageAI, and we report averages for 30 test cases in each scenario. Bold entries are the best-performing. Arrows also mean the same as in the above table.}
\label{tbl:main_llm}
\centering
\resizebox{0.98\textwidth}{!}{%
\begin{tabular}{llcccccccccccc}
\toprule

\multirow{2}{*}{LLM} & \multirow{2}{*}{Methods} & \multicolumn{6}{c}{Emotion JSD $\downarrow$} & \multicolumn{6}{c}{Cluster Matching (\%) $\uparrow$} \\ \cmidrule(lr){3-8} \cmidrule(lr){9-14} 
 &  & Russophobia & US2020 & COVID & UK & US2024 & Gaza & Rus (72) & US20 (82) & COV (87) & UK (60) & US24 (79) & Gaza (75) \\ \midrule
 
\multirow{2}{*}{Gemma2-9B} & Fewshot & 0.247 & 0.227 & 0.341 & 0.243 & 0.259 & 0.306 & 59.63 & 66.67 & 56.67 & 44.00 & 69.26 & 41.00 \\ 
 & \model & \textbf{0.220} & \textbf{0.225} & \textbf{0.290} & \textbf{0.207} & \textbf{0.252} & \textbf{0.234} & \textbf{64.81} & \textbf{71.11} & \textbf{60.44} & \textbf{45.00} & \textbf{70.74} & \textbf{44.33} \\ \midrule 
\multirow{2}{*}{Qwen2.5-32B} & Fewshot & 0.370 & \textbf{0.273} & 0.380 & 0.308 & 0.441 & 0.352 & \textbf{63.70} & 60.37 & 62.78 & 46.00 & 68.52 & 43.00 \\
 & \model & \textbf{0.305} & 0.287 & \textbf{0.362} & \textbf{0.301} & \textbf{0.357} & \textbf{0.352} & 58.15 & \textbf{64.07} & \textbf{67.22} & \textbf{47.92} & \textbf{70.37} & \textbf{44.67} \\ \midrule
\multirow{2}{*}{Mistral-Large} & Fewshot & 0.228 & 0.221 & 0.320 & 0.236 & 0.306 & 0.261 & 51.85 & 64.81 & 55.56 & 48.67 & 69.26 & \textbf{47.33} \\
 & \model & \textbf{0.195} & \textbf{0.220} & \textbf{0.297} & \textbf{0.201} & \textbf{0.269} & \textbf{0.252} & \textbf{62.59} & \textbf{71.85} & \textbf{61.11} & \textbf{48.67} & \textbf{72.59} & 44.33 \\ \midrule
\multirow{2}{*}{GPT-4o-mini} & Fewshot & 0.326 & 0.289 & 0.360 & 0.287 & 0.376 & 0.393 & 57.41 & 66.30 & \textbf{63.89} & 47.33 & 65.93 & 44.67 \\
 & \model & \textbf{0.294} & \textbf{0.258} & \textbf{0.321} & \textbf{0.244} & \textbf{0.374} & \textbf{0.387} & \textbf{63.33} & \textbf{70.00} & 60.00 & \textbf{48.00} & \textbf{68.15} & \textbf{49.67} \\
 
\midrule \midrule 

\multirow{2}{*}{LLM} & \multirow{2}{*}{Methods} & \multicolumn{6}{c}{LLM Discrimination Score $\uparrow$} & \multicolumn{6}{c}{Cluster Coverage (\%) $\uparrow$} \\ \cmidrule(lr){3-8} \cmidrule(lr){9-14} 
 &  & Russophobia & US2020 & COVID & UK & US2024 & Gaza & Russophobia & US2020 & COVID & UK & US2024 & Gaza \\ \midrule

\multirow{2}{*}{Gemma2-9B} & Fewshot & \textbf{8.167} & 8.100 & 6.994 & 7.947 & 7.722 & \textbf{8.020} & 66.67 & 45.37 & 28.33 & 62.33 & 46.48 & \textbf{60.19} \\ 
 & \model & 7.948 & \textbf{8.337} & \textbf{7.117} & \textbf{8.120} & \textbf{7.989} & 7.927 & \textbf{68.52} & \textbf{53.70} & \textbf{40.00} & \textbf{69.00} & \textbf{50.56} & 55.83 \\ \midrule 
\multirow{2}{*}{Qwen2.5-32B} & Fewshot & 7.944 & \textbf{8.304} & 7.044 & 8.000 & 7.848 & 7.927 & 57.41 & 54.63 & 27.38 & 61.50 & 45.00 & 59.17 \\
 & \model & \textbf{8.022} & 7.900 & \textbf{7.611} & \textbf{8.057} & \textbf{7.941} & \textbf{7.990} & \textbf{68.33} & \textbf{62.04} & \textbf{28.33} & \textbf{64.81} & \textbf{54.26} & \textbf{61.67} \\ \midrule
\multirow{2}{*}{Mistral-Large} & Fewshot & 8.156 & 8.311 & 7.206 & 8.433 & 8.081 & \textbf{8.040} & 74.07 & 63.89 & 26.19 & \textbf{69.83} & 50.19 & 57.50 \\
 & \model & \textbf{8.393} & \textbf{8.515} & \textbf{7.756} & \textbf{8.487} & \textbf{8.089} & 7.930 & \textbf{75.93} & \textbf{69.44} & \textbf{34.35} & 63.17 & \textbf{52.04} & \textbf{70.37} \\ \midrule
\multirow{2}{*}{GPT-4o-mini} & Fewshot & 7.956 & 8.237 & 7.267 & 8.407 & 8.081 & 8.033 & 57.41 & \textbf{62.04} & 28.33 & 58.17 & 48.70 & 48.33 \\
 & \model & \textbf{8.119} & \textbf{8.326} & \textbf{7.511} & \textbf{8.447} & \textbf{8.133} & \textbf{8.053} & \textbf{61.11} & 55.56 & \textbf{31.67} & \textbf{60.67} & \textbf{51.67} & \textbf{60.83} \\

\bottomrule
\end{tabular}
}
\end{table*}

\subsubsection{Datasets}
Below are descriptions of the collected data:
\begin{itemize}
    \item {\bf Russophobia}: Collected from January to December 2022 using pro-Russia and pro-Ukraine keywords in the API filter. Notably, it includes the usage of the term ``Russophobia,'' which some view as a propaganda term promoted by the Russian government. This dataset also incorporates samples from the public dataset~\cite{chen2023tweets}.
    \item {\bf US Election 2020 (US2020)}: Collected from December 2020 to January 2021 using election keywords and candidates' names in the API filters. This dataset also includes samples from the public dataset~\cite{chen2022election2020}.
    \item {\bf COVID}: Collected from January to September 2020, during the onset of COVID. This dataset is sampled from a public dataset~\cite{chen2020tracking}.
    \item {\bf Former UK Prime Minister Liz Truss (UK)}: Collected from June to November 2022 using the former UK Prime Minister's name ``Liz Truss'' as the keyword filter. It contains posts on the former UK Prime Minister before her election and after her resignation.
    \item \textcolor{black}{{\bf US Election 2024 (US2024)}: Sampled 30 popular testing posts from July to December 2024 using keyword filters related to the election, including candidates' names.}
    \item \textcolor{black}{{\bf Israel-Hamas Conflict (Gaza)}: Sampled 30 popular testing posts from October 2023 to January 2024 using keyword filters related to the conflict, such as names of locations and parties involved in the conflict.}
\end{itemize}

\textcolor{black}{For the first four datasets, after reserving 30 posts each for testing, the remaining data are encoded into the historical response database. 
The key difference between the first four scenarios and the last two is whether the historical response module can directly retrieve matching contexts or must rely on broader semantic similarities.
We call the last two adaptability-testing scenarios, which are similar to out-of-distribution tests, highlighting the critical role of the external knowledge retrieval module in transferring knowledge from historical responses and grounding the prediction with external information.}

Due to budget constraints, we sampled 100,000 posts from three public datasets within specific date ranges. Additionally, we retrieved the parent posts if they were parts of cascade structures. We also randomly collected up to 200 replies from the root cascade posts to create more comprehensive datasets. A summary of the dataset statistics can be found in Table~\ref{tbl:datasets}. 

News articles are collected from links mentioned in social media posts and by searching through the GDELT project~\cite{gdelt} using geolocations and keyword filters. Ultimately, we gathered 8,061 news articles from prominent news outlets like CNN and the Washington Post, which are preprocessed according to Section~\ref{1_kg_col_ext}. To obtain valid evaluation results, we mask the historical and external data from the future using timestamps relative to the test posts. In production, databases should only contain data from the past, so this is not a concern.

\subsubsection{Baselines and Metrics}

To demonstrate the flexibility of our \model framework, we tested it with multiple embedding models, including VoyageAI (voyage-3-large), OpenAI (text-embedding-3-large), and the open-source NV-Embed-2 model~\cite{lee2024nv}. We also showcase its performance with various LLMs, including 
Gemma2-9B~\cite{team2024gemma}, 
Qwen2.5-32B~\cite{yang2024qwen2}, 
Llama3.3-70B~\cite{dubey2024llama}, Mistral-Large, and GPT-4o-mini, covering LLMs of different sizes and including a commercial model. We compare our framework against the following baseline methods:
\begin{itemize}
    \item {\bf Direct Prompting}: Prompting the LLM directly with the new post content and instructing it to predict responses with the hint that it is generating a reply on $\mathbb{X}$ platform.
    \item {\bf Few-shot Prompting}: Similar to direct prompting, and at most five response demonstrations from historical responses are added, ordered by their retrieval order.
\end{itemize}

External news snippets are also included in the baseline prompts to ensure a fair comparison, especially in adaptability-testing scenarios. We use VoyageAI to encode articles, conduct regular vector retrieval, and inject them into the prompts.

To quantitatively evaluate the performance of response forecasts, we use the following metrics:
\begin{itemize}
    \item {\bf Emotion JSD}: We let the LLM extract the emotion contents~\cite{plutchik1989measuring} of both predicted and real responses, resulting in two distributions after normalization. We calculate their Jensen-Shannon Divergence (JSD) to assess the emotional similarity of the predicted responses (bounded by 1).
    \item {\bf LLM Discrimination Score}: We prompt the LLM to rate the likelihood of each predicted response being another real one on a scale from 1 to 10 after showing 10 selected examples from real responses, ordered by their popularity and text lengths.
    \item {\bf Cluster Matching Percentage}: We cluster the real responses using the same method described in Section~\ref{3_cluster}. After embedding the predicted responses into the same space, we calculate the proportion of them that belong to one of the clusters (disregarding outliers). Higher matching indicates that the predicted responses are more relevant.
    \item {\bf Cluster Coverage Percentage}: Similar to the previous metric, after clustering real responses, we calculate the percentage of clusters covered by at least one predicted response. Higher coverage suggests that the predicted responses have greater diversity in ideology and semantics.
\end{itemize}
We opted for automatic metrics due to the high volume of generated texts and budget constraints. To ensure more reliable metrics, we average results across Qwen2.5-72B, Llama3.3-70B, and Mistral-Large for metrics requiring an LLM, and across VoyageAI and OpenAI for those needing an embedding model for clustering. Since we disregard outliers for the cluster matching percentage, we report its upper bound alongside the scenario name in the table.

\begin{table}[h]
\centering
\caption{Response Generation Example}
\label{tbl:examples}
\resizebox{\columnwidth}{!}{%
\begin{tabular}{p{0.88\linewidth}}
\toprule
\multicolumn{1}{c}{\textbf{Input Post, $p_s$}} \\ \midrule

``Hamas' military wing publishes scenes showing its members fighting at close range against Israeli occupation forces in different parts of Gaza City. Some with armored vehicles, bulletproof vests, artillery, aviation. The others with a tracksuit, some old sneakers, an RPG" \\

\midrule\midrule
\multicolumn{1}{c}{\textbf{Generated Responses}} \\ \midrule

{\hspace{1.5em}\em -- Supporting Gaza:\/} \\ \midrule

``I see you're sharing some intense footage from Gaza. It's heartbreaking to witness such \textbf{disparity in resources} and the devastation civilians face in this ongoing struggle.'' \\ \midrule

``This just highlights the brutal \textbf{asymmetry of the conflict}. One side has advanced military capabilities, while the other is fighting with sheer willpower and desperation. It's a stark reminder of the tragic reality in Gaza.''  \\ \midrule

``The situation in Gaza is heartbreaking and complex. We must remember \textbf{the human cost behind this conflict}.'' \\ \midrule

{\hspace{1.5em}\em -- Anti-Hamas:\/} \\ \midrule

``Hamas’s tactics might \textbf{seem heroic in their propaganda}, but the sad reality is innocent lives are caught in this crossfire.''  \\ \midrule

``Looks like Hamas is putting on to \textbf{paint a brave picture}, but we all know \textbf{how this started and the reality is much darker.}'' \\ 

\midrule

``It’s disturbing to see this \textbf{glorification of violence}. While Hamas prepares for combat, innocent folks pay the price.'' \\ \midrule

{\hspace{1.5em}\em -- Anti-Israel:\/} \\ \midrule

``Both sides are suffering, yet the way Israel operates raises questions \textbf{about their true agenda}. \#GazaWar \#HumanRights'' \\ \midrule

``I imagine the aversion they have for them, they have \textbf{been mistreated for 75 years}.'' \\ \midrule

``The suffering in Gaza is unimaginable. How much longer will the world \textbf{turn a blind eye}? \#CeasefireNow'' \\

\bottomrule
\end{tabular}%
}
\end{table}

\subsection{Evaluation Results}

Table~\ref{tbl:main_emb} and Table~\ref{tbl:main_llm} present detailed evaluation results for \model using various choices of embedding models and LLMs with a forecast quota of 30. 
In summary, \model achieves an average improvement of 10.2\% in emotion JSD, 1.5\% in LLM discrimination score, 4.1\% in cluster matching percentage, and 11.8\% in cluster coverage percentage compared to the baselines. Table~\ref{tbl:main_emb} specifically demonstrates \model's superiority over baseline methods across all four metrics. It also illustrates the universal applicability of our framework to various embedding models. Table~\ref{tbl:main_llm} further supports these observations with different LLMs, where our framework performs well in all scenarios.
The significant improvements in emotion JSD and cluster coverage demonstrate the high effectiveness of our framework in generating responses that better capture emotion distribution and ideological diversity. Additionally, improvements in cluster matching and discrimination score suggest that our framework can incorporate retrieved information to foster a more relevant and realistic forecast than only invoking the internal knowledge in the LLMs.

\begin{table*}[h!]
\caption{\textcolor{black}{Evaluation results for the ablation study where \model components are taken offline individually. ``Ideo'' means including ideological embeddings for clustering (combined clustering). ``SP'' means sparse retrieval of news articles rather than dense retrieval. Bold entries are the best-performing, and the underlined ones are the second-best-performing.}}
\label{tbl:ablation1}
\centering
\resizebox{0.98\textwidth}{!}{%
\begin{tabular}{lcccccccccccc}
\toprule
\multirow{2}{*}{Methods} & \multicolumn{6}{c}{Emotion JSD $\downarrow$} & \multicolumn{6}{c}{Cluster Matching (\%) $\uparrow$} \\ \cmidrule(lr){2-7} \cmidrule(lr){8-13} 
 & Russophobia & US2020 & COVID & UK & US2024 & Gaza & Rus (72) & US20 (82) & COV (87) & UK (60) & US24 (79) & Gaza (75) \\ \midrule
 
Full (\model) & \textbf{0.223} & \textbf{0.214} & \textbf{0.265} & \textbf{0.223} & \textbf{0.283} & \textbf{0.212} & \textbf{55.56} & \textbf{67.04} & \textbf{58.33} & \textbf{48.67} & \underline{68.89} & \textbf{47.67} \\
 w/o KG & 0.263 & \underline{0.246} & 0.310 & \underline{0.246} & \underline{0.295} & \underline{0.228} & \underline{54.81} & \underline{66.67} & \underline{57.33} & \underline{45.00} & 67.04 & \underline{46.00} \\
 w/o KG \& Ideo & \underline{0.253} & 0.252 & \underline{0.298} & 0.252 & 0.299 & 0.238 & 52.96 & 66.30 & 56.67 & 42.33 & \textbf{72.22} & 43.67 \\
 w/o KG \& Ideo \& SP & 0.270 & 0.275 & 0.305 & 0.262 & 0.318 & 0.234 & 48.15 & 62.22 & 56.67 & 43.67 & 64.07 & 41.00 \\
 
\midrule\midrule

\multirow{2}{*}{Methods} & \multicolumn{6}{c}{LLM Discrimination Score $\uparrow$} & \multicolumn{6}{c}{Cluster Coverage (\%) $\uparrow$} \\ \cmidrule(lr){2-7} \cmidrule(lr){8-13} 
 & Russophobia & US2020 & COVID & UK & US2024 & Gaza & Russophobia & US2020 & COVID & UK & US2024 & Gaza \\ \midrule
 
Full (\model) & \textbf{8.552} & \textbf{8.730} & \textbf{7.350} & \textbf{8.967} & \textbf{8.163} & \textbf{8.140} & \textbf{77.78} & \underline{63.89} & \textbf{37.50} & \textbf{66.50} & \underline{60.19} & \textbf{62.50} \\
 w/o KG & \underline{8.515} & 8.570 & \underline{7.244} & \underline{8.830} & \underline{8.130} & \underline{8.117} & \underline{75.93} & \textbf{68.52} & \underline{36.67} & \underline{65.17} & \textbf{61.30} & \underline{57.50} \\
 w/o KG \& Ideo & 8.433 & \underline{8.656} & 7.228 & 8.717 & 8.096 & 8.107 & 70.37 & 55.56 & 33.33 & 63.17 & 49.81 & 54.17 \\
 w/o KG \& Ideo \& SP & 8.419 & 8.644 & 7.150 & 8.770 & 8.067 & 8.097 & 62.96 & 58.33 & 25.00 & 62.96 & 44.26 & 54.17 \\
 
\bottomrule
\end{tabular}
}
\end{table*}
\begin{table*}[h!]
\caption{Evaluation results for the ablation study on two design choices for \model. Bold entries are the best-performing and underlined ones are the second-best-performing.}
\label{tbl:ablation2}
\centering
\resizebox{0.98\textwidth}{!}{%
\begin{tabular}{lcccccccccccc}
\toprule

\multirow{2}{*}{Methods} & \multicolumn{3}{c}{Emotion JSD $\downarrow$} & \multicolumn{3}{c}{LLM Discrimination Score $\uparrow$} & \multicolumn{3}{c}{Cluster Matching (\%) $\uparrow$} & \multicolumn{3}{c}{Cluster Coverage (\%) $\uparrow$} \\ \cmidrule(lr){2-4} \cmidrule(lr){5-7} \cmidrule(lr){8-10} \cmidrule(lr){11-13} 
 & Russophobia & UK & US2024 & Russophobia & UK & US2024 & Rus (72) & UK (60) & US24 (79) & Russophobia & UK & US2024 \\ \midrule
 
Full (\model) & \textbf{0.223} & \textbf{0.223} & \textbf{0.283} & \textbf{8.552} & \textbf{8.967} & \textbf{8.163} & \textbf{55.56} & \textbf{48.67} & \underline{68.89} & \textbf{77.78} & \textbf{66.50} & \textbf{60.19} \\
 w/o UMAP & \underline{0.264} & 0.245 & \underline{0.317} & 8.326 & 8.763 & 8.133 & \underline{54.81} & \underline{47.33} & \textbf{70.00} & \underline{72.22} & \underline{63.17} & \underline{59.07} \\
 w/o Inst & 0.273 & \underline{0.239} & 0.345 & \underline{8.544} & 8.700 & \underline{8.152} & 51.85 & 46.00 & 68.52 & \underline{72.22} & 62.96 & 54.26 \\
 w/o Inst \& UMAP & 0.284 & 0.246 & 0.350 & 8.230 & \underline{8.810} & 8.126 & 51.33 & 45.00 & 65.93 & 64.81 & 62.33 & 49.81 \\
 
\bottomrule
\end{tabular}
}
\end{table*}

\subsection{Response Generation Example}
We present a concrete response generation example from the adaptability-testing scenario (Israel-Hamas Conflict) in Table~\ref{tbl:examples} to show the effectiveness of our framework. The predicted responses successfully capture diverse ideological perspectives with emotions.
Quantitative evaluation results on emotion JSD and discrimination score align with these qualitative observations, showing high relevance to the input post. The semantic coherence, realism, and ideological diversity of the predicted responses confirm the capability of our framework.

\subsection{Ablation Study}

We conducted ablation studies on the importance of each component in our system. We use VoyageAI and Llama3.3 in all these experiments. In Table~\ref{tbl:ablation1}, we observe consistently improved performance when all components are present, with a gradual decline as they are taken offline, indicating that they work together to achieve optimal results. The second-best-performing variant is \model without KG, demonstrating that knowledge graph relations, which are often more generic, can further inform the generation process alongside news articles. When the ideological embedding and sparse retrieval components are disabled, performance drops to slightly above the baseline methods. This indicates that ideological information is essential for the community identification and that the sparse retrieval of news articles is a better choice in our setting where input posts are typically short texts.

In Table~\ref{tbl:ablation2}, we demonstrate the effectiveness of two design choices: UMAP dimension reduction of semantic embedding and a special instruction for embedding models. \textcolor{black}{Due to space constraints, we only display results for three datasets, but the conclusion remains the same.} In most cases, the variant with UMAP outperforms the one without it, and a similar conclusion can be drawn regarding the special embedding instruction. This outcome is expected, as ideological embedding is weakened without UMAP, given that its dimensionality is significantly less than that of raw semantic embedding. The special instruction is essential because the emphasis should be on the response of the augmented posts rather than disregarding their structure.

\section{Related Work}\label{sec:related}

Predicting responses on social media is challenging due to its informal and dynamic nature. The data-driven approach~\cite{ritter2011data} pioneered response generation in social media by utilizing large datasets to learn response patterns. To improve relevance, TA-Seq2Seq~\cite{xing2017topic}, a topic-aware neural response generation model, incorporates topic information. Another approach to response generation is context-aware prototype editing~\cite{wu2019response}, where models generate responses by editing existing prototypes based on the conversational context. Finally, CGRG~\cite{wu2021controllable} has developed a controllable model for grounded response generation, allowing for more precise control over responses. While previous work has explored information retrieval techniques, context-aware generation, and factual grounding, they have not investigated the potential of LLMs. Their applications are typically limited to autocompletion or chatbots, whereas we aim to develop a framework that predicts community responses, enabling various social applications.

RAG integrates retrieval mechanisms with generative models to improve knowledge-intensive tasks. REALM~\cite{guu2020retrieval} introduced retrieval-augmented language model pre-training, leveraging external knowledge to improve language understanding. This method is further developed in \cite{lewis2020retrieval} for knowledge-intensive NLP tasks, integrating retrieval directly into the generation process. There are surveys~\cite{li2022survey,gao2023retrieval} that offer comprehensive overviews of RAG methodologies and their applications. 
Despite these applications, standard RAG methods face challenges in highly dynamic environments like social media, primarily due to the computational expense. As a result, the historical response database is updated infrequently. 
In contrast to dense retrieval methods, sparse retrieval techniques such as TF-IDF and BM25~\cite{robertson2009probabilistic} are less computationally expensive but depend on term matching, which leads to lower accuracy. 
SPLADE~\cite{formal2021splade} improves retrieval accuracy by employing query expansion techniques to address the term matching issue, making it suitable for our goal of searching longer news articles based on shorter social media posts. 
Inspired by social computing, our framework utilizes a combined RAG approach of dense and sparse retrieval to meet different needs.

Various representation learning techniques have been developed to encode social entities. The interactions between users and posts encapsulate their beliefs and preferences, which can be leveraged to extract ideological embeddings for both users and posts. Existing research often models the interaction history as graphs to derive social representations. The variational graph autoencoders (VGAE)~\cite{kipf2016variational} introduce a framework featuring a GCN-based encoder and an inner-product decoder that maps users and messages to a latent space with normally distributed variables. InfoVGAE~\cite{li2022unsupervised} and SGVGAE~\cite{li2024large} propose non-negative VGAE models capable of capturing the ideologies of users and messages, encoding them into an interpretable space. We utilize these techniques to enhance semantic embeddings and clustering.

\section{Conclusion}\label{sec:conclusion}

In this paper, we presented \model, a framework that leverages social computing-based RAG methods integrated with LLMs to predict community responses to real or hypothetical social media posts. Through an extensive experimental evaluation across six real-world scenarios on the $\mathbb{X}$ platform, we demonstrated that \model consistently outperforms baseline methods, generating realistic, diverse, and contextually grounded forecasts. Further experiments using various embedding models and LLMs confirmed the modularity and robustness of \model. A response generation example illustrated the framework's effectiveness at accurately capturing ideological and emotional nuances reflective of real-world community dynamics. Ablation studies highlighted the contribution of each component within \model, emphasizing the importance of the community-aware historical response module and the external knowledge module for achieving optimal performance. Overall, \model provides a powerful tool for applications in social computing such as public sentiment forecasting and social what-if analysis.

Moving forward, we plan to extend our framework by adding multimodal capabilities, integrating multimedia data alongside textual inputs to further improve prediction accuracy, which will provide even deeper insights into evolving social conversations.

\section*{Acknowledgments}
Research reported in this paper was sponsored in part by
DEVCOM ARL under Cooperative Agreement W911NF-172-
0196, NSF CNS 20-38817, and the Boeing Company. It was
also supported in part by ACE, one of the seven centers
in JUMP 2.0, a Semiconductor Research Corporation (SRC)
program sponsored by DARPA. The views and conclusions
contained in this document are those of the authors and
should not be interpreted as representing the official policies,
either expressed or implied, of the Army Research Laboratory,
or the U.S. Government. The U.S. Government is
authorized to reproduce and distribute reprints for Government
purposes notwithstanding any copyright notation herein.

\bibliographystyle{IEEEtran}
\bibliography{ref}

\end{document}